\documentclass[twocolumn]{revtex4-1}
\usepackage{graphicx}
\usepackage{epstopdf}
\usepackage[lofdepth,lotdepth]{subfig}
\usepackage{amsmath}
\usepackage{caption}

\begin{document}

\title{Room temperature deposition of superconducting Niobium Nitride films by ion beam assisted sputtering}

\author{Tomas Polakovic}
\affiliation{Physics Division, Argonne National Laboratory, Argonne, IL~60439, USA}
\affiliation{Department of Physics, Drexel University, Philadelphia, PA~19104, USA}
\author{Sergi Lendinez}
\author{John E.~Pearson}
\author{Axel Hoffmann}
\affiliation{Materials Science Division, Argonne National Laboratory, Argonne, IL~60439, USA}
\author{Volodymyr Yefremenko}
\affiliation{High Energy Physics Division, Argonne National Laboratory, Argonne, IL~60439, USA}
\author{Clarence L.~Chang}
\affiliation{High Energy Physics Division, Argonne National Laboratory, Argonne, IL~60439, USA}
\author{Whitney Armstrong}
\affiliation{Physics Division, Argonne National Laboratory, Argonne, IL~60439, USA}
\author{Kawtar Hafidi}
\affiliation{Physics Division, Argonne National Laboratory, Argonne, IL~60439, USA}
\author{Goran Karapetrov}
\affiliation{Department of Physics and Department of Materials Science and Engineering, Drexel University, Philadelphia, PA~19104, USA}
\author{Valentine Novosad}
\email{Corresponding author: novosad@anl.gov}
\affiliation{Materials Science Division, Argonne National Laboratory, Argonne, IL~60439, USA}
\affiliation{Physics Division, Argonne National Laboratory, Argonne, IL~60439, USA}

\begin{abstract}
We use room temperature ion beam assisted  sputtering (IBAS) to deposit niobium nitride thin films. Electrical and structural characterizations were performed by electric transport and magnetization measurements at variable temperatures, X-ray diffraction and atomic force microscopy. Compared to reactive sputtering of NbN, films sputtered in presence of an ion beam show remarkable increase in the superconducting critical temperature T$_{\rm{c}}$, while exhibiting lower sensitivity to nitrogen concentration during deposition. Thickness dependence of the superconducting critical temperature is comparable to films prepared by conventional methods at high substrate temperatures and is consistent with behavior driven by quantum size effects or weak localization.
\end{abstract}

\maketitle

\section{Introduction}
Niobium nitride (NbN) has long been a material of interest for fabrication of nano-devices in the field of quantum electronics, such as superconducting nanowire single photon detectors\cite{Goltsman} and superconducting SIS tunnel junctions\cite{kawakami2001fabrication} due to its relatively high superconducting T$_{\rm{c}}$, large superconducting energy gap and ease of fabrication. However, one of the undesired features of NbN is the presence of multiple crystal structure modifications, not all of them superconducting.\cite{oya1974transition} Conventional fabrication methods, such as reactive sputtering\cite{shoji1992superconducting} or CVD and thermal diffusion\cite{gurvitch1985tunneling}, stabilize the growth of a desired phase of NbN by carrying out the deposition at elevated temperatures, generally more than 500\,$^{\circ}$C, which makes the process incompatible with methods like lift-off, heterostructure growth with materials sensitive to heat or fabrication of tunnel junctions, where impurity diffusion leads to interaction at the junction interface. Recently the superconductivity with high critical temperature in hard $\epsilon$-NbN grown at high pressure and high temperatures was discovered.\cite{zou_scirep2016} While there exist processes capable of achieving high-T$_{\rm{c}}$ NbN films deposited at room temperature, such as incorporation of methane gas with RF diode sputtering\cite{cukauskas1983effects}, they lead to films with a granular or columnar void structure, resulting in normal state resistivity well above 10$^4$\,$\mu \Omega\cdot\rm{cm}$\cite{cukauskas1983effects,kawakami2001fabrication}. Use of substrate biasing during deposition procedure has also been demonstrated to be a viable method of room temperature deposition, but deposited films show presence of lower-T$_{\rm{c}}$ tetragonal phase of NbN, which leads to suppression of overall film critical temperature.\cite{dane2017bias} These growth methods require extremely precise control over relative concentrations of sputtering gasses and large sputtering powers, which leads to substrate heating.\cite{bacon1983properties}

Use of ion beam bombardment during deposition process is known to have dramatic impact on the microstructure of films.\cite{smidt1989ion, harper1984ion} It leads to densification of films and increases adhesion\cite{hirsch1980thin} and the additional kinetic energy supplied by the ion beam allows for increased mobility of the atomic species near the surface, reducing the presence of voids that are substituted by dislocation boundaries\cite{hibbs1984effects, marchenko2008film}. Also, the increase in the momentum anisotropy leads to development of texture with preferred orientation of film grains\cite{muller1985dependence, dobrev1982ion, van1980crystalline, ma2004texture}. In this work we explore ion beam assisted sputtering (IBAS) that combines N$_2$ bombardment with conventional DC magnetron sputtering. We show that IBAS can be used to produce NbN thin films with superior superconducting properties even with deposition carried out with substrate at room temperatures. We demonstrate that use of neutralized nitrogen ion beam during DC magnetron sputtering from a Nb target leads to NbN films with relatively high superconducting transition temperatures of up to 14.5\,K and with normal resistivity as low as 110.62\,$\mu\Omega\cdot\rm{cm}$, without any need for substrate heating or biasing. A direct comparison of IBAS to conventional DC reactive sputtering carried out in the same chamber shows not only an increase in superconducting critical temperature T$_{\rm{c}}$ of NbN films, but also a large decrease in process sensitivity to nitrogen concentration, leading to more consistent results compatible with large scale fabrication. We conduct structural and electrical characterization of the IBAS-grown non-epitaxial thin NbN films focusing on the mechanism of suppression of superconductivity in very thin films grown on Si wafers. We find that thin films grown by IBAS have good superconducting properties down to a critical thickness of approximately 2\,nm, which is comparable to films grown by conventional methods on epitaxial substrates at high temperatures. We demonstrate that the evolution of superconducting transition temperature with film thickness T$_{\rm{c}}(d)$ can be explained by quantum size effects or, potentially, weak localization.

\section{Sample Fabrication Methods}
NbN films were prepared by DC magnetron sputtering in a commercial ultra-high vacuum sputtering system  (Angstrom Engineering).\cite{Angstrom} After transferring the silicon wafer substrate with thermally grown silicon oxide through a load-lock, the chamber was pumped down to less than $5\times$10$^{-8}$\,Torr before commencing the deposition procedure. Before the actual sputtering step, the substrate's surface was treated using a low energy argon ion beam. This is done primarily to eliminate any water or organic contamination and without loss of more than 1\,nm of the substrate surface.

Sputtering was carried out at 2\,mTorr with Ar$_2$ (99.9999\% purity) as sputtering gas. During reactive sputtering, N$_2$ (99.9997\% purity) gas was mixed into the sputtering gas, while in ion beam assisted sputtering, nitrogen was supplied through the ion gun only. The amount of argon and nitrogen was controlled by mass flow controllers and monitoring of residual and sputtering gasses was done by a quadrupole gas analyzer, to ensure equivalent gas mixture in the chamber for comparison of reactive and ion beam assisted sputtering.

The 3\,in dia. sputtering target consisted of 99.9999\% pure Nb, it was located 5\,in. away from the substrate at a 33$^{\circ}$ angle relative to substrate surface normal and powered at 0.18\,kW from a DC magnetron power source. Sputtering rates of approximately 1\,\AA/s were determined from a calibrated quartz thickness monitor and confirmed after deposition by X-Ray reflectometry or by profilometer measurement on a shadowmasked twin sample. A slight difference in sputtering rates is observed when compared to reactive sputtering, which is approximately 15\% lower under equivalent conditions.

The ion beam source was an End-Hall ion gun\cite{kaufman1987end}, in which neutralization of nitrogen ions was achieved by thermionic emission of electrons from a hollow cathode. The ion gun, positioned at azimuthal 20$^{\circ}$ relative to the sputtering gun and at an angle of 40$^{\circ}$ relative to sample surface normal, was operated in constant gas flow mode, to facilitate comparison with reactive sputtering, while discharge and emission currents and voltages were kept constant during deposition. As ion bombardment at energies above 300\,eV are known to cause structural damage to thin films\cite{kern2012thin}, the energy of the ion beam was kept at relatively low value of 100\,eV per N$_2$ in order to minimize these effects. This translates to ion beam power densities of approximately 70\,mW/cm$^2$. Total ion beam current under these conditions was maintained at nominal 0.5\,A. The value of 100 eV per N$_2$ was deemed optimal based on comparison with depositions carried out at 50, 200 and 300\,eV, which lead to maximum T$_{\rm{c}}$ of 12.9, 13.6 and 13.1\,K, respectively. The reduction of T$_{\rm{c}}$ at lower energies can be explained by decrease in mobility of adatoms and defects due to reduction of available kinetic energy from the incoming ions.

The films were grown on polished Si substrates with native oxide without any intentional heating during deposition. Self-heating due to sputtering did not exceed 55$^{\circ}$C, as determined by a calibrated thermocouple built into the substrate holder assembly.

\section{Results and Discussion}
\subsection{Characterization of thick NbN films}
\begin{figure}
	\centerline{\includegraphics [clip, width=\linewidth, angle=0]{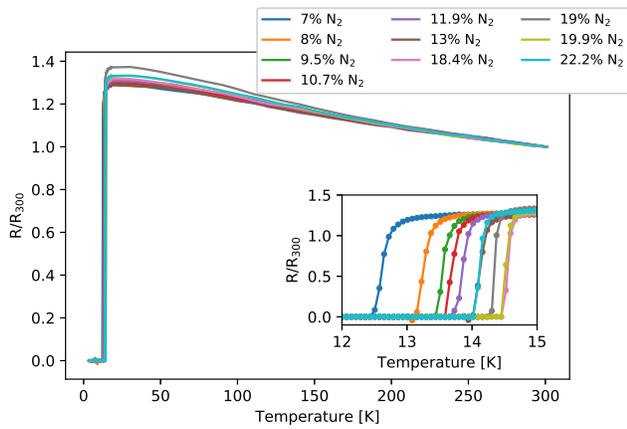}}
	\caption[justification=raggedright,singlelinecheck=false]{Normalized resistance as a function of temperature for 240\,nm thick films grown by ion beam assisted sputtering at various N$_2$ concentrations.}
	\label{nitrogen:RT}
\end{figure}

\begin{figure}
	\centerline{\includegraphics [clip,width=\linewidth, angle=0]{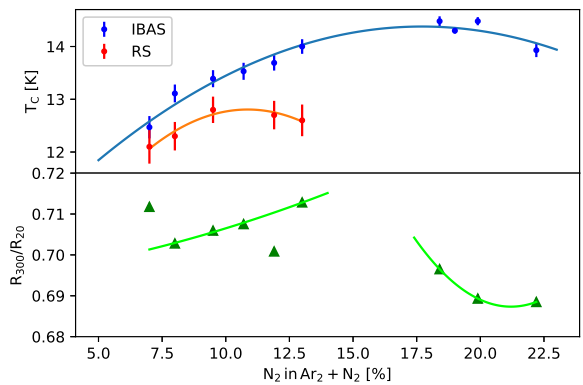}}
	\caption{Dependence of superconducting T$_{\rm{c}}$ (top) of 240 nm thick films on nitrogen concentration. Blue points correspond to ion beam assisted sputtering (IBAS), red points correspond to reactive sputtering. Error bars denote the 90\% - 10\% transition width. Residual resistance ration (green, bottom) is for 240\,nm IBAS films. Trend lines are polynomial fits meant as guides for the eye.}
	\label{nitrogen:Cdep}
\end{figure}

The superconducting $\rm{T}_{\rm{c}}$ and residual resistivity ratio were measured using the standard four probe technique in a Quantum Design PPMS. Resistive transitions of 240\,nm thick NbN films deposited on Si substrate by IBAS can be seen Fig.~\ref{nitrogen:RT}. To better demonstrate the dependence of superconducting $\rm{T}_{\rm{c}}$ on the concentration of nitrogen and to facilitate comparison to reactive sputtering, the superconducting transition temperatures are explicitly plotted as a function of nitrogen concentration in Fig.~\ref{nitrogen:Cdep}, with the general trend of both curves that follows results in literature\cite{kawakami2001fabrication, oya1974transition, shoji1992superconducting, cukauskas1983effects, bacon1983properties}: the superconducting $\rm{T}_{\rm{c}}$ peaks and the transition width shrinks as the NbN film approaches optimal stoichiometry. However, one can clearly see a quantitative difference when comparing the two techniques. First, there is a significant difference in the highest value of the superconducting $\rm{T}_{\rm{c}}$, with the IBAS samples reaching 14.5\,K, close to optimum value for bulk stoichiometric NbN. Second, there is obvious decrease in the process sensitivity towards the concentration of nitrogen in the growth chamber: one can achieve $\rm{T}_{\rm{c}}>$\,14\,K in a range of concentrations from 13\% to 22\% - a dramatic improvement from reactive sputtering, in which high superconducting T$_{\rm{c}}$ is constrained to a window of approximately 2\%.\cite{kawakami2001fabrication} Room-temperature resistivity of the thin films was  110.6$\pm$6.6\,$\mu\Omega\cdot\rm{cm}$, showing no noticeable trend with N$_2$ concentration.

\begin{figure}
	\centerline{\includegraphics [clip,width=\linewidth, angle=0]{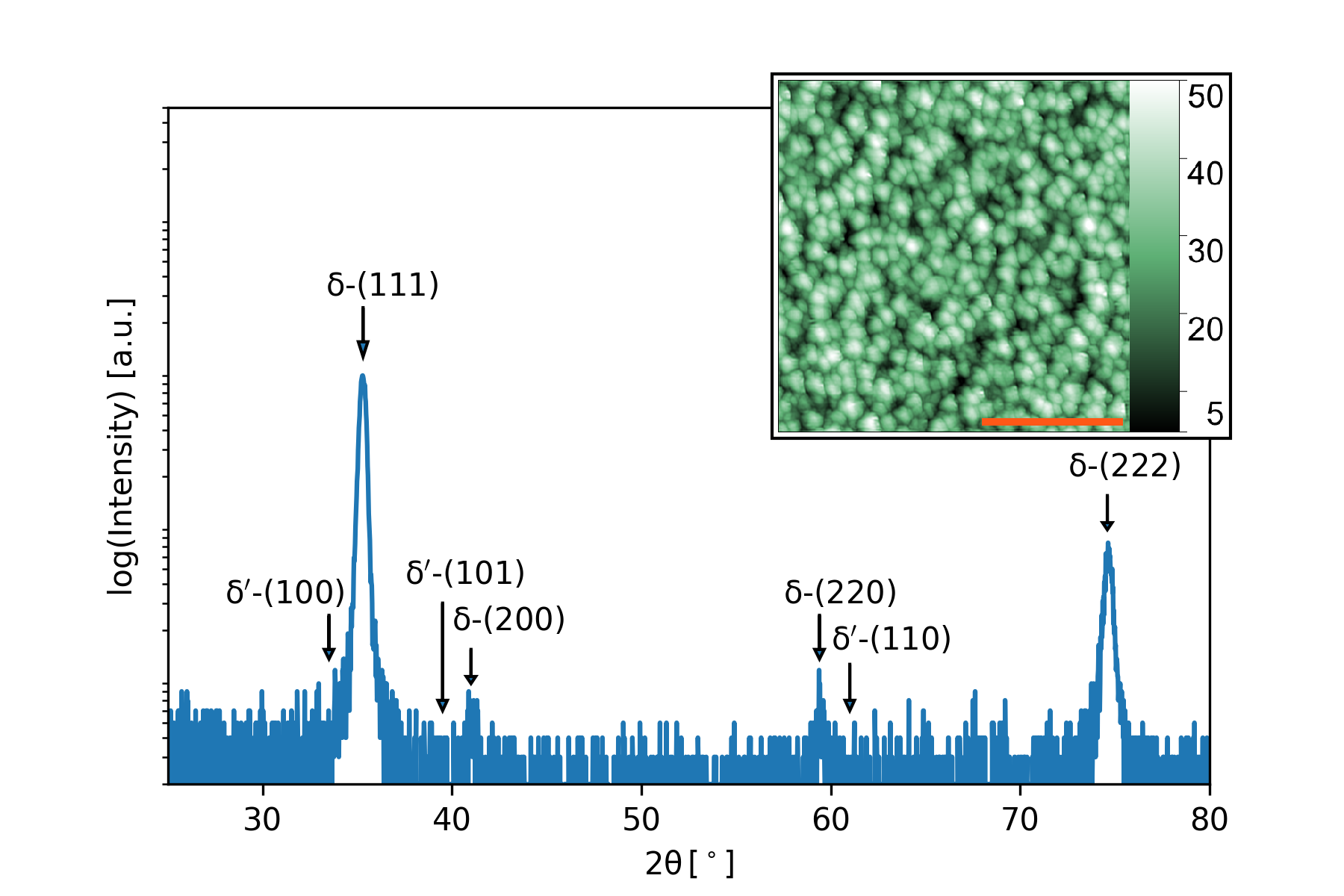}}
	\caption{X-ray diffraction pattern of a 500 nm NbN film deposited at optimal conditions on Si substrate. All visible diffraction peaks correspond to cubic $\delta$-NbN. Inset: AFM scan of film surface. Scale bar corresponds to 400 nm.}
	\label{XRD}
\end{figure}

We confirmed close to optimal stoichiometry for the phase with the highest T$_{\rm{c}}$ using X-ray diffraction (Fig.~\ref{XRD}). We observe prominent peaks of the cubic $\delta$-NbN, without any presence of the non-superconducting phases, such as the common $\delta^\prime$-NbN phase.\cite{benkahoul2004structural}

Despite the results indicating textured films containing predominantly cubic $\delta$-NbN phase, the superconducting T$_{\rm{c}}$ of the films is lower than that of single crystal NbN. This could be explained by effects of grain boundaries suppressing the local density of states, leading to reduced total T$_{\rm{c}}$, even if the intragrain T$_{\rm{c}}$ would be close to maximum.\cite{tyan1994grain,nigro1988electrical} This effect was observed in some of our magnetization measurements, where the superconducting transition has a long tail of more than 1~Kelvin. This is further corroborated by the residual resistivity ratios RRR = $R_{300}/R_{20}$, which are all smaller than unity, an effect attributed to grain boundary scattering of conduction electrons.\cite{tyan1994grain} Consistently with this description, the RRR correlates with the superconducting T$_{\rm{c}}$, reaching a maximum value of approximately 0.72 (Fig.~\ref{nitrogen:Cdep}). By correlating this value with findings in literature, this RRR corresponds to average grain size of approximately 25\,nm\cite{nigro1988electrical}, in agreement with values we have determined by XRD (22\,nm) and AFM (mean grain width of 25 $\pm$ 5 nm) measurements. Additionally, the limited presence of voids in films deposited by IBAS might come at the expense of increased density of dislocation defects\cite{marchenko2008film}. This micro-structural disorder can cause additional electron scattering, increasing the film resistance, decreasing the RRR and also potentially lead to weak localization as discussed in the next section.

To determine upper critical magnetic field and coherence length, we carried out magnetization measurements at various fields close to superconducting T$_{\rm{c}}$, where the temperature dependent H$_{\rm{c2}}$(T) was defined as a field at which the magnetization vanishes. The upper critical field H$_{\rm{c2}}$(T=0\,K) was calculated by extrapolation from the Werthamer-Helfand-Hohenberg formula\cite{werthamer1966fields}:

\begin{equation}
	H_{\rm{c2}}(0) = -0.69 \;  T_{\rm{c}} \left. \frac{dH_{\rm{c2}}(T)}{dT} \right|_{T_{\rm{c}}}.
\end{equation}

The  in-plane coherence length was obtained from the Ginzburg-Landau theory, where~\cite{Tinkham1996superconductivity}:

\begin{equation}
	H_{\rm{c2}}(T) = \frac{\Phi _0}{2 \pi \xi^2 (T) }.
	\label{eq:xi}
\end{equation}

\begin{figure}
	\centerline{\includegraphics [clip,width=\linewidth, angle=0]{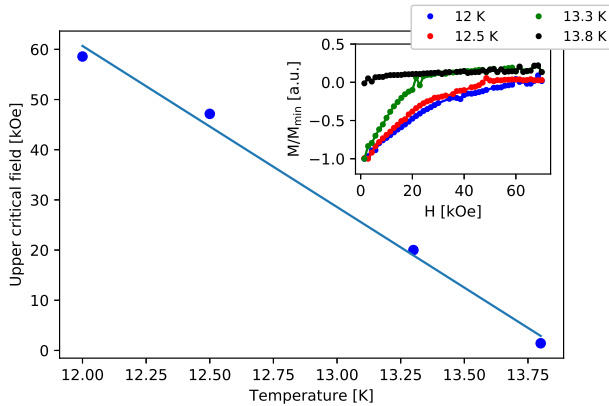}}
	\caption{Perpendicular upper critical field H$_{\rm{c2}}$ measured as a function of temperature for a 240 nm thin film deposited at optimal conditions. Inset: Normalized magnetization of the same film as a function of applied field at temperatures close to T$_{\rm{c}}$.}
	\label{Hc2}
\end{figure}

Strictly speaking, this dependence should be valid only in the critical region close to superconducting T$_{\rm{c}}$, but in practice, it can be applied even deep into the superconducting state. From the upper critical fields measured for film grown at optimal conditions (Fig. \ref{Hc2}), the extrapolated perpendicular critical field was determined to be H$_{\rm{c2}}$(0) = 319\,kOe and the estimated coherence length is $\xi(0)$ = 3.2\,nm, slightly smaller than the bulk value of 5\,nm reported in literature\cite{jha2012fabrication, vasyutin2016upper}. This reduced value of $\xi(0)$ is the result of the renormalization of coherence length due to short electron mean-free path in disordered sputtered films.\cite{faucher2002niobium}

\subsection{Suppression of superconductivity in ultra-thin films}

As many applications of superconducting devices necessitate for the material to be in a form of a thin film, we also study the dependence of superconducting and electronic properties as a function of film thickness. It is well known that in NbN the superconducting state is suppressed as the film becomes thin,\cite{Ilin2003suppresion, Engel2006frontiers, semenov2009optical} and it is usually explained either by weak localization\cite{graybeal1984localization, xiao1996proximity, kim2006proximity}, electron wave leakage\cite{yu1976consistent} or surface contribution to the Ginzburg-Landau free energy of the superconductor.\cite{simonin1986surface}

\begin{figure}
	\includegraphics [clip,width=\linewidth, angle=0]{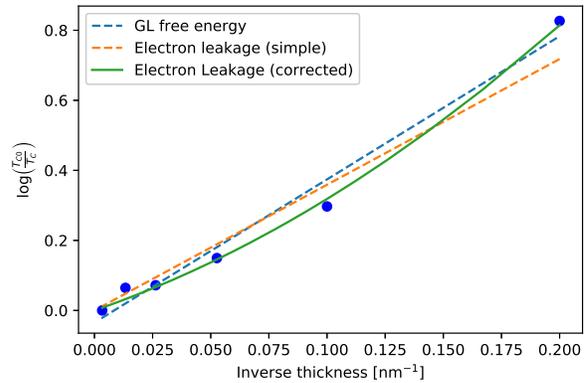}
	\caption{Dependence of superconducting T$_{\rm{c}}$ on inverse film thickness. Experimental data is plotted in blue circles and lines show best fits of different models: Green solid line is a fit of Eq. (\ref{eq:leakage_corr}), dashed orange corresponds to fit of Eq. (\ref{eq:leakage_simple}) and blue dashed line is a fit of Eq. (\ref{eq:McMillan}).}
	\label{T_thin:TvsInvD}
\end{figure}

One way to determine which model best fits our experimental data is to look at the dependence of the superconducting T$_{\rm{c}}$ on film thickness (Fig.~\ref{T_thin:TvsInvD}). In electron leakage model, the the electron wave function is considered to be quantized in the direction perpendicular to sample surface. This quantization leads to reduction in density of states and allows for the wave function to leak outside of the superconductor. The simplified theory predicts a behavior of superconducting T$_{\rm{c}}$ as\cite{yu1976consistent, kang2011suppression}:

\begin{equation}
	\frac{T_{\rm{c}}}{T_{\rm{c}\infty}} = \exp \left[ \frac{-b}{N(0)Vd} \right],
	\label{eq:leakage_simple}
\end{equation}

where $T_{\rm{c}\infty}$ is the critical temperature of bulk, $b$ is the characteristic length of electron wave leakage, approximately equal to the electron Fermi wavelength, and $N(0)V$ is the BCS coupling. If we assume $N(0)V$ = 0.32\cite{kang2011suppression}, the estimated $b$ = 1.14\,\AA{} is reasonably close to the reported values for NbN. However, considering the disordered nature of sputtered films, one might want to use a version of Eq. (\ref{eq:leakage_simple}) corrected for presence of defects and film breakup:

\begin{equation}
	\frac{T_{\rm{c}}}{T_{\rm{c}\infty}} = \exp \left[ \frac{-1}{N(0)V} \left( \frac{b}{d} + \frac{c}{d^2} \right) \right],
	\label{eq:leakage_corr}
\end{equation}

where $c$ is a term describing contribution of defects and is typically in the range from 0 to 20\,\AA$^2$. Usage of parameters reported on previous films \cite{kang2011suppression} leads to quantitative behavior similar to the uncorrected theory. Removing this restriction allows for a quantitatively better fit, with estimated values $b$ = 0.73\,\AA{} and $c$ = 2.84\,\AA$^2$. The length of $b$ is not significantly shorter than the reported values and $c$ falls within expected range, meaning that the estimate is not unphysical. The difference from values reported by Kang~\textit{et. al} might be explained by difference in the microstructure of our films, as evidenced by different sheet resistance of thin films produced by our IBAS method.

Considering the approximately linear trend of superconducting T$_{\rm{c}}(d)$, a variational result from modified Ginzburg-Landau theory with an added surface term could also be applied\cite{simonin1986surface}:

\begin{equation}
	\frac{T_{\rm{c}}}{T_{\rm{c}\infty}} = \left[ 1 - \frac{2a}{N (0) V} \frac{1}{d} \right],
	\label{eq:GL}
\end{equation}

where $a$ is the Thomas-Fermi screening length. Using this model, we can extrapolate the limiting thickness where the superconducting state vanishes as ${\displaystyle d_m=2.7~\rm{nm}}$, which is comparable to the coherence length $\xi(0)$ extracted from Eq.~(\ref{eq:xi}) and supports the notion that ion beam assisted sputtering achieves growth without considerable amount of non-superconducting interfacial layers, even on substrates with considerable lattice mismatch. Further, we can estimate the value of screening length ${a\approx0.4\,\rm{nm}}$,~in good agreement with the assumption of it being on the order of lattice spacing\cite{chockalingam2009tunneling,piatti2016superconducting} and much smaller than the coherence length $\xi(0)$.

More insight into the behavior of the superconducting state in thin films can be gained from the dependence of T$_{\rm{c}}$ on the films' sheet resistance R$_{\rm{sheet}}$. Ivry~\textit{et. al.} proposed a phenomenological power-law dependence of the form\cite{ivry2014universal}:

\begin{equation}
	d\cdot T_{\rm{c}} = A R_{\rm{sheet}}^{-B},
	\label{Ivry1}
\end{equation}

where $d$ is the film thickness and $A$ and $B$ are fitting constants. This equation can be rewritten into a form:

\begin{equation}
    T_{\rm{c}} = \left(\frac{A}{d} \right) \exp \left[-B \ln{ \left( R_{\rm{sheet}} \right)}\right],
	\label{Ivry2}
\end{equation}

which can be contrasted to the result derived from BCS theory\cite{mcmillan1968transition}:

 \begin{equation}
 	T_{\rm{c}} = \frac{\Theta_{\rm{D}}}{1.45} \exp \left[ -\frac{1.04 (1 + \lambda)}{ \lambda - \mu (1 + 0.62 \lambda)} \right],
	\label{eq:McMillan}
\end{equation}

where $\Theta_{\rm{D}}$ is the Debye temeparture, $\lambda$ is the electron-phonon coupling constant and $\mu$ describes Coulomb repulsive interactions. When comparing equations (\ref{Ivry2}) and (\ref{eq:McMillan}), one can see that the term $B$ is related to changes in the BCS coupling $N(0)V$ (in weak coupling limit equal to $\lambda - \mu$) or the interaction parameters $\lambda$ and $\mu$, which would scale as $B\ln ( R_{\rm{sheet}} )$. Fitting our experimental data to equation (\ref{Ivry1}) yields a result $B=2$, different from the average value of $B\sim1$ (although still within range of reported values\cite{ivry2014universal}).

\begin{figure}
	\includegraphics [clip,width=\linewidth, angle=0]{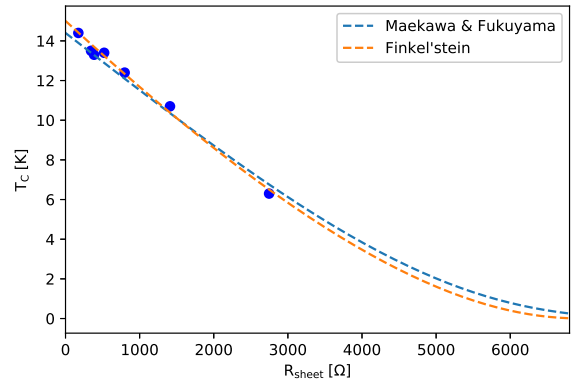}
	\caption{Dependence of superconducting T$_{\rm{c}}$ on sheet resistance. Blue curve is a fit of the experimental data to Eq. (\ref{eq:pertubation}), orange curve corresponds to fit to Eq. (\ref{eq:finkel})}
	\label{T_thin:TvsRs}
\end{figure}

A more quantitative approach to study of Coulomb interactions and localization effects can be also applied. Perturbation theory for localization in 2D superconductors, developed by Maekawa and Fukuyama, yields a result for superconducting T$_{\rm{c}}$ in the form\cite{maekawa1982localization}:

\begin{equation}
\begin{split}
	\ln{\left( \frac{T_{\rm{c}}}{T_{\rm{c}\infty}} \right)} = &-\frac{1}{2} \frac{g_1 N(0) e^2 R_{\rm{sheet}}}{2 \pi^2 \hbar} \left[ \ln{\left(5.5 \frac{\xi}{l} \frac{T_{\rm{c}\infty}}{T_{\rm{c}}} \right)} \right]^2 \\
	&-\frac{1}{3} \frac{g_1 N(0) e^2 R_{\rm{sheet}}}{2 \pi^2 \hbar} \left[ \ln{\left(5.5 \frac{\xi}{l} \frac{T_{\rm{c}\infty}}{T_{\rm{c}}} \right)} \right]^3,
\end{split}
\label{eq:pertubation}
\end{equation}

where $\xi$ is the coherence length, $l$ is the electronic mean free path and $g_1 N(0)$ is an effective BCS coupling constant. In the dirty limit of $\xi \gg l$, the first term, which is due to reduction of density of states, becomes negligible when compared to the second term, corresponding to a vertex correction to the electron-electron interaction. Under these assumptions, the superconducting T$_{\rm{c}}$ reduction should have an approximately linear dependence on the film sheet resistance. While our results do show linear behavior (as seen in Fig. \ref{T_thin:TvsRs}), a fit to Eq. (\ref{eq:pertubation}) provides  the effective BCS coupling constant $g_1$N(0)\,=\,23.62, which is an unphysically large correction to the standard BCS value of 0.32. Also, a superconductor to insulator transition can typically be driven by weak localization when the sheet resistances $R_{\rm{sheet}}$ are around the quantum value \(\displaystyle\frac{h}{4e^2} \approx\) 6.4 $k\Omega$~\cite{kagawa1996superconductor}, as can be seen also when extending our fit. Extrapolating the dependence of R$_{\rm{sheet}}$ vs film thickness, the critical thickness for this transition is approximately 2\,nm, which is close to the estimate from Eq.~(\ref{eq:GL}).
Alternatively, one can employ Finkel'stein's results using renormalization group methods\cite{finkel1996superconducting,*finkelstein_jetplett1987,*finkelstein_physicab_1994}:

\begin{equation}
	\displaystyle\frac{T_{\rm{c}}}{T_{\rm{c}\infty}} = \exp \left[ \frac{-1}{\gamma} \right] \times \left[ \frac{1 + \frac{\sqrt{r / 2}}{\gamma - r / 4}}{1 - \frac{\sqrt{r / 2}}{\gamma - r / 4}} \right] ^{1/\sqrt{2d}},
	\label{eq:finkel}
\end{equation}

where $\displaystyle\gamma = \frac{1}{\log (k_B T_{\rm{c}\infty}\tau / \hbar)}$, $\displaystyle r=\frac{R_{\rm{sheet}}}{(2 \pi ^2 \hbar / e^2)}$, k$_{\rm{B}}$ is the Boltzmann constant, $e$ is the elementary charge and $\tau$ is the electron  elastic scattering time. Fitting our data to this equation yields $\tau$ = 2.43$\times10^{-15}$\,s, four times smaller than the value reported in literature\cite{ezaki2012localization}. This difference is not surprising when one considers higher resistance of our thin films and its relation to the electronic mean-free path, which is proportional to $\tau$.

Both equations (\ref{eq:pertubation}) and (\ref{eq:finkel}) predict vanishing superconductivity at values of R$_{\rm{sheet}}$ that coincide with critical thicknesses extrapolated from models related to dimensionality effects (equations (\ref{eq:leakage_simple}), (\ref{eq:leakage_corr}) and (\ref{eq:GL})), which complicates determination of suppression mechanism. Multiple results in literature observe critical thickness for NbN close to 2\,nm in films prepared using different growth conditions and therefore having different electronic properties\cite{semenov2009optical, kang2011suppression, wang1996superconducting, ezaki2012localization, makise2015superconductor, ivry2014universal}, which might be an indication that the reduction of superconducting T$_{\rm{c}}$ is driven by effects related to dimensionality. However, there has been no observation of higher order effects, such as T$_{\rm{c}}$ oscillation with thickness predicted by the electron leakage model, in this work or others. As Eq.\,(\ref{eq:finkel}) yields a good quantitative fit with physically reasonable values, further study is required to rule out localization effects as a mechanism for suppression of T$_{\rm{c}}$, even more so, when one considers the disordered nature of sputtered films.

\section{Conclusions}
We have shown that NbN thin film growth using low energy bombardment with N$_2$ during sputtering of Nb has beneficial effects on superconducting and electronic properties of the resulting NbN films. This room temperature process results in films having resistivity as low as 110\,$\mu\Omega$.cm, relatively high T$_{\rm{c}}$ of 14.5\,K and critical magnetic field $H_{\rm{c2}}(0)$ of nearly 32\,T. The stoichiometric growth can be achieved in a broad range of nitrogen concentrations and does not require epitaxial growth conditions, which opens opportunities for broader application of NbN in quantum electronic devices.

Our data on ultrathin NbN films supports the predictions of models of electron wave leakage (quantum size effect) or weak localization. Even on non-epitaxial substrates, superconductivity persists down to thickness of approximately 2\,nm, which also coincides with sheet resistances equal to resistance quantum, where one can potentially expect superconductor-insulator transition driven by weak localization.\\

\section{Acknowgledments}
The authors would like to thank Aaron Miller from Quantum Opus, LLC in Novi, MI USA and Andr{\'e} Anders from Leibniz Institute of Surface Engineering in Leipzig, Germany for stimulating discussion.

This work was supported by the U. S. Department of Energy (DOE), Office of Science, Offices of Nuclear Physics, Basic Energy Sciences, Materials Sciences and Engineering Division under Contract \# DE-AC02-06CH11357. G.K. was supported by the Center for the Computational Design of Functional Layered Materials (CCDM), an Energy Frontier Research Center funded by the U.S. Department of Energy, Office of Science, Basic Energy Sciences under Award \# DE-SC0012575.


%

\end{document}